\documentstyle [12pt] {article}
\begin{document}
\thispagestyle{empty}
\begin{flushright} DESY 98-104\\August 1998\
\end{flushright}
\mbox{}
\vspace{0.5in}
\begin{center}
{\Large \bf Degenerate Dirac Neutrinos\\[0pt]}
\vspace{1.0in}
{\sc Utpal Sarkar$^{(a,b)}$}\footnote{E-mail: utpal@prl.ernet.in}  

\vskip 0.8cm

\begin{small} 
$^{(a)}$Theory Group, DESY, Notkestra{\ss}e 85, 22607 Hamburg, Germany\\
$^{(b)}$Theory Group, Physical Research Laboratory,
Ahmedabad, 380 009, India \footnote{permanent address}\\
\end{small}
\end{center}

\vskip 2cm

\begin{abstract}

A simple extension of the standard model is proposed in which all
the three generations of neutrinos are Dirac particles and
are naturally light. We then assume that the neutrino mass matrix
is diagonal and degenerate, with a few eV mass
to solve the dark matter problem. The self energy 
radiative corrections, however,
remove this degeneracy and allow mixing of these neutrinos.
The electroweak radiative corrections then predict a 
lower bound on the $\nu_\mu - \nu_e$
mass difference which solves the solar neutrino problem through MSW 
mechanism and also predict a lower bound on the 
$\nu_\tau - \nu_\mu$ mass difference which is just enough to 
explain the atmospheric neutrino problem as reported by super Kamiokande. 

\end{abstract}

\newpage 
\baselineskip 18pt 

The neutrinoless double beta decay \cite{double}
puts severe constraints on the Majorana
mass of the $\nu_e$. With this constraint it is not possible to explain 
simultaneously the 
dark matter problem \cite{dm}, solar neutrino problem \cite{sol}, 
atmospheric neutrino problem \cite{atm}
and the laboratory bounds on the mixing angles \cite{lab}. 
Since the preliminary results from KARMEN \cite{karmen}
contradicts the LSND result \cite{lsnd}, 
we shall not include that in our analysis. If
the neutrinos are Dirac particle there will not be any lepton number
violation and hence there will not be any constraint from the neutrinoless
double beta decay. In that case one can postulate an almost degenerate 
neutrino scenario \cite{almost} to explain the other problems. 

In this article, we consider exactly degenerate Dirac 
particles with a few eV mass to explain
the dark matter problem. However, this will not allow any flavour 
mixing. So we introduce explicit lepton flavour violation, which will
break the mass degeneracy radiatively, which in turn will allow 
flavour mixing. The electroweak self-energy corrections will then 
predict a lower bound on the 
mass squared differences between $\nu_\mu$ and $\nu_e$
which can solve the solar neutrino problem through matter enhanced
neutrino oscillation \cite{msw}
and simultaneously predict a lower bound on the 
mass squared difference between $\nu_\tau$ and  $\nu_\mu$ which 
is just enough to solve the atmospheric neutrino problem. There is also
a similar contribution to mass splitting from flavour violating 
radiative corrections, which also gives the maximal neutrino flavour mixing.

Consider a two generation Majorana neutrino 
scenario. The neutrino mass matrix is given 
by, \begin{equation}  M_\nu = \pmatrix{ m_{ee} & m_{e \mu} \cr m_{\mu e}
& m_{\mu \mu}}. \end{equation}
The neutrinoless double beta decay \cite{double}
will imply $m_{ee} < 0.46$ eV. If we consider an almost degenerate 
neutrino scenario \cite{almost}, to solve the dark matter problem \cite{dm} 
we require $m_{\mu \mu} \simeq m_{ee}$. Then the small mixing \cite{lab}
of $\nu_e$ with $\nu_\mu$ will imply that $sin^2 \theta_{e \mu}
\sim {m_{e \mu} \over m_{\mu \mu}} < 0.6$ and the masses of the $\nu_e$ and
$\nu_\mu$ are less than 1 eV. This will not solve the dark matter
problem. This problem is solved if one assumes that
the neutrinos are Dirac particles. However, the main problem of making
the neutrinos a Dirac particle is that, 
in a simple extension of the standard 
model the Dirac mass of the neutrinos are related to the charged lepton 
masses and hence cannot be small (of the order of a few eV) naturally. 

We now propose a scenario where the neutrinos are 
Dirac particles and are naturally light. 
The left handed neutrinos combine with the right handed 
neutrinos through their interactions with a different higgs doublet,
which does not couple to the quarks and charged leptons because of the 
presence of an additional $U(1)$ symmetry. This new higgs doublet 
acquires a small vacuum expectation value ($vev$), when the extra $U(1)$
symmetry and the electroweak symmetry are broken \cite{ma},
and hence its coupling gives
a small Dirac mass to the neutrinos naturally. There is no lepton number
violation in this scenario and hence there are no Majorana mass of the
neutrinos. 

We extend the standard model gauge group to include a new $U(1)$ 
symmetry, $${\cal G}_{ext} \equiv 
SU(3)_c \times SU(2)_L \times U(1)_Y \times U(1)_X. $$ 
We also extend the model to include the three right handed neutrinos
$\nu_{iR}$ ($i=1,2,3$); four additional singlet fermions $Y_i$ ($i=1,2,3$)
and $Z$, which are required for purpose of anomaly cancellation;
one new higgs doublet $\eta$, a scalar singlet $\chi$ and a charged
scalar singlet $\zeta$. Transformation
properties of the new particles are presented in table 1. 
\begin{table}
\caption{transformations of the new particles}
\center
\begin{tabular}{||l|c||}
\hline \hline
& Fermions \\
\hline
$\nu_{i R}$ &(1,1,0,2) \\
\hline
$Y_{i}$ & (1,1,0,-1) \\
\hline
$Z$ & (1,1,0,-3) \\
\hline \hline
&Scalars \\
\hline
$\eta$& (1,2,-1/2,-2) \\
\hline
$\chi$ & (1,1,0,2) \\
\hline
$\zeta$ & (1,1,1,-2) \\
\hline \hline
\end{tabular}
\end{table}
The scalar
$\chi$ acquires a $vev$ at a very high scale M, breaks the $U(1)_X$
symmetry and give masses to the extra singlet fields. The mass of the 
doublet $\eta$ is also of the order of M,
but it does not acquire any $vev$ to start with. However, 
after the electroweak symmetry breaking it acquires a small $vev$ and
gives small Dirac masses to the neutrinos naturally.

The lagrangian contains the quadratic and the quartic terms and the 
trilinear mixing terms which are given by,
\begin{eqnarray}
{\cal L}_{scalar} &=& M^2_\eta \eta^\dagger \eta + 
M^2_\chi \chi^\dagger \chi
+ m^2_\phi \phi^\dagger \phi \nonumber \\
&+& {1 \over 2} \lambda_1 (\phi^\dagger \phi)^2 +
{1 \over 2} \lambda_2 (\eta^\dagger \eta)^2 +
{1 \over 2} \lambda_3 (\chi^\dagger \chi)^2 \nonumber \\ &+&
{1 \over 2} \lambda_4 (\phi^\dagger \phi)(\eta^\dagger \eta) +
{1 \over 2} \lambda_5 (\chi^\dagger \chi)(\eta^\dagger \eta) +
{1 \over 2} \lambda_6 (\phi^\dagger \phi)(\chi^\dagger \chi) \nonumber \\ 
&+& \mu \phi^\dagger \eta \chi .
\end{eqnarray}
The scalar $\chi$ acquires a $vev$ at some large scale $M$, which is
the only other mass scale in the model other than the electroweak 
symmetry breaking scale. We consider 
$$M_\chi \sim M_\eta \sim <\chi> \sim M \sim 10^{11} GeV. $$
Then to prevent the usual higgs doublet from acquiring a large mass
we require $$\mu \sim m_\phi \sim m , $$ where $m$ is the electroweak
syummetry breaking scale. 

We may now minimize the potential to determine the $vev$s of the different
scalar fields. They are given by,
\begin{eqnarray} 
<\chi>^2 \sim - {M_\chi^2 \over \lambda_3}; ~~~~~<\phi>^2 \sim
{- \lambda_1 m_\phi^2 + \lambda_4 M_\eta^2
\over \lambda_1 \lambda_2 - \lambda_4^2} \nonumber \\
{\rm and}~~~~~
<\eta> \sim {\mu <\phi> <\chi> \over M_\eta^2} \sim 100 eV.
\end{eqnarray}
The scalar field $\eta$ acquires a very small $vev$ naturally in this
scenario. As a result, if it gives a Dirac mass to the neutrinos, then
we have a natural explanation of the smallness of the Dirac mass of the 
neutrinos. To obtain the mass of the neutrinos we now write down the 
Yukawa couplings of the leptons and the singlet fermions,
\begin{equation}
{\cal L} = f_{\nu i \alpha} \overline{l_{iL}} \nu_{\alpha R} \eta + g_{ab} 
\overline{Y_a^c} Y_b \chi + g^\prime_a \overline{Y_a} Z \chi 
+ f_{e i \alpha} \overline{l_{iL}} e_{\alpha R} \phi + h_{\alpha \beta}
\overline{(\nu_{\alpha R})^c} e_{\beta R} \zeta .
\end{equation}
The extra singlets $Y_a$ and $Z$ get very large masses from the $vev$
of the scalar $\chi$ and they donot couple with the light neutrinos. 
Since they are decoupled from the low energy sector, we shall
not discuss them from now on. Although we have to introduce these singlet
fields for purpose of anomaly cancellation, if one can implement this
mechanism in a larger theory like grand unified theory or string theory, 
one may not require to introduce these fields. The scalar $\zeta$ 
give radiative mass splitting to neutrinos of different generations, 
which in turn gives neutrino flavour mixing. 

The first term in this equation gives small Dirac mass to the neutrinos
\begin{equation}
M_\nu = f_{\nu i \alpha} <\eta>  .
\end{equation}
We now assume that the Dirac mass matrix is diagonal and degenerate 
and to explain the hot component of the dark matter the diagonal 
elements are given by $f_{\nu i \alpha} \delta_{i \alpha} 
= m$. The neutrino Dirac mass matrix then becomes 
$$ M_\nu = \pmatrix{ m & 0 & 0 \cr 0 & m & 0 \cr 0& 0& m} .$$
We further assume that in this basis the Yukawa interactions of the 
field $\zeta$ is also diagonal, but the
charge lepton mass matrix is not diagonal. In general, it may be 
possible to diagonalise the charge lepton mass matrix and the
neutrino mass matrix simultaneously, since the neutrino mass 
matrix is degenerate. However, the coupling of the charged scalar $\zeta$
will give radiative correction, which will break the 
mass degeneracy and will not allow to make the charged current interaction
diagonal. As a result all the observed
mixing may come from the charge lepton mass matrix and we can only determine
them from experiments. The mixing angle will depend on the amount
of radiative mass splitting. 

In general, the charged lepton mass matrix can be diagonalised by a 
bi-unitary transformation $$ V_{ik}^\dagger M_{e i \alpha} U_{\alpha \beta} =
M^{diag}_{e k \beta} \delta_{k \beta} .$$ 
Then the matrix $V_{ik}$, which diagonalises the matrix $M_{e i \alpha}
M^\dagger_{e j \alpha} $, will enter in the charged current interactions.
In the basis $[e_{iL},~e_{\alpha R}]$, in which the charged lepton mass 
matrix is diagonal, the charge current 
interaction of the neutrinos and the charged leptons will be given
by, \begin{equation}  {\cal L}_{cc} = \overline{\nu_{i L}} \gamma_\mu
V^\dagger_{k i} e_{k L} W^\mu . \end{equation} 
The charged lepton mixing matrix
$V_{ij}$ will introduce neutrino flavour mixing after the radiative 
mass splitting.

We now define the basis for the neutrinos $[\nu^e_{iL},~\nu^e_{\alpha R}]$,
which has diagonal charged current interaction and are given by,
$$ \nu^e_{i L} = V_{i k} \nu_{k L} ~~~~~ {\rm and} ~~~~~ 
\nu^e_{\alpha R} = U^\nu_{\alpha \beta} \nu_{\beta R}. $$
In this basis, the mass matrix is not diagonal. But when the mass
matrix is diagonal and degenerate, we can always make transformations
to the right handed fields and make them diagonal. In the basis
$[\nu^e_{iL},~\nu^e_{\alpha R}]$, the neutrino mass matrix is given
by, $$  M^e_{\nu i j} = V_{ ij } M_{\nu }  . $$ 
But $V_{ij}$ commutes with $M_\nu$ and hence we can make a
transformation $\nu^e_{\alpha R} = V_{\beta \alpha}^\dagger \nu_{\beta R};
~~~ V_{\beta \alpha} = V_{ij}$, and diagonalise the mass matrix.
However, in the presence of the radiative corrections due to $\zeta$,
this is not possible. 

We assume that the couplings of $\zeta$ to the right handed leptons,
$h_{\alpha \beta}$, to be diagonal. Although the phenomenology of such
dilepton have not been studied, one can extend the analysis of ref.
\cite{dilepton} to constrain the parameters. If, in addition, we assume
that $h_{11} < 10^{-5}$, then there is only one constraint from the
$(g-2)_\mu$, which is $h_{22} > 0.3$ for $m_\zeta \sim 100$ GeV. For
$h_{33}$ there is no bound in this scenario and we can consider this
to be of the order of 1. 
With this choice we get a self energy radiative correction
(with the internal loop containing charged leptons and $\zeta$)
to the neutrino mass matrix in the basis
$[\nu_{iL},~\nu_{\alpha R}]$, in which the charged current interaction
is not diagonal but the $\zeta$ couplings are diagonal, given by
\begin{equation}
M_{\nu i \alpha} = \pmatrix{m + m^\zeta_1 & 0 & 0 \cr 0 & m + m^\zeta_2
& 0 \cr 0 & 0 & m + m^\zeta_3}
\end{equation}
where, $m^\zeta_i = m {h_{ii}^2 \over 4 \pi }{e_i^2 \over m_\zeta^2}$;
$e_i = e, \mu, \tau$. In this case, it will not be possible to diagonalise
the neutrino mass matrix in the basis in which the charged current interaction
is diagonal. Including the standard model self energy 
radiative corrections \cite{rad},
we can now write down the neutrino mass matrix in the basis
$[\nu^e_{iL},~\nu^e_{\alpha R}]$ as follows,
\begin{equation}
M^e_{\nu i \alpha} = V_{ij} M_{\nu i \alpha} + 
{\rm diag}[m^{ew}_1 ~~m^{ew}_2 ~~m^{ew}_3 ] ~ V_{ ij } M_{\nu j \alpha}
\end{equation}
where, $m^{ew}_i = \alpha_w 
(m_{e_i}^2 / m_w^2) $,~ $e_i \equiv e, \mu, \tau$.
This mass matrix can now be diagonalised to get the flavour mixing
matrix and the mass ssquared difference between the different
flavours of neutrinos. 
For any arbitrary choice of the mixing matrix it is difficult 
to solve this analytically. So, for purpose of illustration we 
demonstrate with a two generation example, and then present some 
realistic numbers for a three generation scenario which we check 
numerically.

We consider the $\mu$ and $\tau$ family in the two generation example.
For $V_{ij} = \pmatrix{ \cos ~\theta & - \sin ~\theta \cr 
\sin ~\theta & \cos ~\theta}$, the mass matrix is given by,
\begin{equation}
M^e_{\nu i \alpha} = \pmatrix {
(m + m^\zeta_2)(1 + m^{ew}_2) \cos ~\theta
& - (m + m^\zeta_3)(1 + m^{ew}_2) \sin ~\theta \cr
(m + m^\zeta_2)(1 + m^{ew}_3) \sin ~\theta &
(m + m^\zeta_3)(1 + m^{ew}_3) \cos ~\theta }
\end{equation}
which can be diagonalised with a bi-unitary transformation. The 
unitary matrix, which diagonalises $M^e_{\nu i \alpha} 
(M^e_{\nu j \alpha})^\dagger$, gives the neutrino flavour mixing
and is given by,
$$ V^e_{ij} = \pmatrix{ \cos ~\phi &\sin ~\phi \cr \sin ~\phi &\cos ~\phi}
$$  where, $$ \phi \sim {1\over 2} \tan^{-1} \left[ 2 ~\cos \theta
~\sin \theta~ {(m^\zeta_3 - m^\zeta_2) \over 
(m^\zeta_3 - m^\zeta_2) + (m^{ew}_3 - m^{ew}_2) } \right] $$
and the mass squared difference is given by,
\begin{equation}
(m_3^2 - m_2^2)^2 = [2 m \cos \theta
\sin \theta (m^\zeta_3 - m^\zeta_2) ]^2 + 
[m (m^\zeta_3 - m^\zeta_2) + m (m^{ew}_3 - m^{ew}_2)]^2 .
\end{equation}
It is clear from the expressions that both the electroweak 
radiative correction as well as the radiative corrections due to $\zeta$
will contribute to the mass squared difference. If $\zeta$ does not 
contribute to the mass difference, the mixing angle vanishes as pointed 
out earlier. For the mixing angle to be maximal we require the 
contribution of $\zeta$ to be of the same order or more than the 
contribution from the electroweak radiative corrections. 

The mass difference generated by the standard 
model self energy radiative corrections
(considering the Dirac mass of the neutrinos to be around 7 eV)
are given by,
\begin{eqnarray}
\Delta m_{sol} &=& 
(m^{ew}_2)^2 - (m^{ew}_1)^2 = \alpha_w m^2 
{m_\mu^2 - m_e^2 \over m_w^2} = 0.6 \times 10^{-5}  \nonumber \\
\Delta m_{atm} &=& (m^{ew}_3)^2 - (m^{ew}_2)^2 = \alpha_w m^2
{m_\tau^2 - m_\mu^2 \over m_w^2} = 
1.4 \times 10^{-3} 
\end{eqnarray}
The $\nu_e - \nu_\mu$ mass difference $\Delta m^{ew}_{sol}$ is just enough
to solve the solar neutrino problem, while the $\nu_\mu - \nu_\tau$ 
mass difference $\Delta m^{ew}_{atm}$ falls within the solution suggested 
by the recent super kamiokande result. If the contribution due to 
$\zeta$ are of the same order of magnitude (which is the case for 
$m_\zeta \sim 100$ GeV and $h_{11} = 0$, $h_{22} \sim 0.1$ and $
h_{33} \sim 1$), then we get the maximal 
mixing angle for the $\nu_\mu$ and $\nu_\tau$ oscillations and the
relevant mass squared difference as required by the super kamiokande
experiment. For the solar neutrino problem, the mass squared difference
between $\nu_e$ and $\nu_\mu$
is just right for the MSW solution \cite{msw}, but we only get the
small mixing solution.
In this scenario it will be difficult to explain
the vacuum oscillation solution of the solar neutrino problem, since
although one can adjust the parameters of the couplings of $\zeta$ to
get a small radiative corrections, one cannot change the electroweak 
radiative correction. As a result, the mass squared difference between
$\nu_e$ and any other neutrinos will be larger than required, unless
there are new sterile neutrinos, to which the $\nu_e$ oscillates. 
Since the neutrino 
masses arise from a different higgs doublet, the neutrino mass mixing 
is not related to the CKM quark mixing matrix. 

In summary, we pointed out that if neutrinos are Dirac particles, then
we may start with a degenerate diagonal mass matrix with the diagonal 
elements to be around a few eV, so that neutrinos could be the hot dark 
matter of the universe. We also presented a model in which neutrinos 
could be light Dirac particles naturally. The mass degeneracy is broken 
by self energy radiative corrections,
which can then allow enough mixing of these neutrinos to solve the
atmospheric neutrino and solar neutrino problems. The standard model
radiative corrections give a lower bound on the mass squared difference
which is just enough to solve both the atmospheric neutrino and solar
neutrino problems. 

\vskip 0.75in
\begin{center} {ACKNOWLEDGEMENT}
\end{center}

I would like to thank Profs. David Cline, Prof. D.P. Roy
and M. Raidal for discussions. I would also like to thank
Prof W. Buchmuller and the Theory Division, DESY,
Hamburg for hospitality. Financial support from the
Alexander von Humboldt Foundation is acknowledged.

\newpage
\baselineskip 16pt
\bibliographystyle{unsrt}

\end{document}